\documentclass{elsart}

 \usepackage{graphicx}
\usepackage{amssymb}

\begin{document}

\begin{frontmatter}



\title{Phase transitions of a tethered membrane model on a torus with intrinsic curvature}
\author[label1]{Isao Endo} and
\author[label2]{Hiroshi Koibuchi}
\ead{koibuchi@mech.ibaraki-ct.ac.jp}

\address[label1]{Department of Electrical and Electronic System Engineering, Ibaraki College of Technology, 
Nakane 866, Hitachinaka,  Ibaraki 312-8508, Japan}

\address[label2]{Department of Mechanical and Systems Engineering, Ibaraki College of Technology, 
Nakane 866, Hitachinaka,  Ibaraki 312-8508, Japan}

\begin{abstract}
%Text of abstract
A tethered surface model is investigated by using the canonical Monte Carlo simulation technique on a torus with an intrinsic curvature. We find that the model undergoes a first-order phase transition between the smooth phase and the crumpled one. 
\end{abstract}

\begin{keyword}
Phase Transition \sep Intrinsic Curvature \sep Elastic Membranes 
\PACS  64.60.-i \sep 68.60.-p \sep 87.16.Dg
\end{keyword}
\end{frontmatter}

% main text
%\section{}
%\label{}
%----------------------------------------------------------
\section{Introduction}\label{intro}
%----------------------------------------------------------
Recently, it has been growing that the concern with elastic surface models of Helfrich and Polyakov-Kleinert \cite{HELFRICH-NF-1973,POLYAKOV-NPB-1986,Kleinert-PLB-1986,David-TDQGRS-1989,NELSON-SMMS-2004}. A considerable number of studies have been conducted on the phase transition between the smooth phase and the crumpled one over the past two decades \cite{DavidGuitter-EPL-1988,Peliti-Leibler-PRL-1985,BKS-PLA-2000,BK-PRB-2001,Kleinert-EPJB-1999,BOWICK-TRAVESSET-PREP-2001,WIESE-PTCP19-2000,WHEATER-JP-1994,CATTERALL-NPB-SUPL-1991,AMBJORN-NPB-1993,BCHGM-NPB-1993,ABGFHHE-PLB-1993,KOIB-PLA-2002,KOIB-PLA-2004,KOIB-EPJB-2005,KOIB-PLA-2003-2,KOIB-PRE-2003,KOIB-EPJB-2004,KOIB-PLA-2005-1,BCFTA-1996-1997,WHEATER-NPB-1996,KANTOR-NELSON-PRA-1987,GOMPPER-KROLL-PRE-1995,KOIB-PRE-2004-1,KOIB-PRE-2004-2,KOIB-PRE-2005-1,KANTOR-SMMS-2004,KANTOR-KARDER-NELSON-PRA-1987,BCTT-EPJE-2001,BOWICK-SMMS-2004,NELSON-SMMS-2004-2}.  

Curvature energies play a crucial role in smoothing the surface. According to curvature energies,  surface models can be divided into two classes; one with an extrinsic curvature and the other with an intrinsic curvature. It is also possible that both extrinsic and intrinsic curvatures are included in a model Hamiltonian. Intrinsic curvature is the one that is defined only by using the metric tensor (first fundamental form) of the surface, and extrinsic curvature is defined by using the extrinsic curvature tensor  (second fundamental form) of the surface \cite{FDAVID-SMMS-2004-2}. Both of the mean curvature $H$ and the Gaussian curvature $K$ defined by using the extrinsic curvature tensor, are considered as an extrinsic curvature. However, the Gaussian curvature can also be considered as an intrinsic curvature, because of the relation $2K\!=\!R$, where $R$ is the scalar curvature defined only by using the metric tensor. 

Intrinsic curvature models were first studied by Baillie et. al. in \cite{BJ-PRD-1993-1994,BEJ-PLB-1993,BIJJ-PLB-1994,FW-PLB-1993}. The shape of surfaces can be strongly influenced by intrinsic curvatures. Recently several numerical studies have been made on the phase diagram of the model with intrinsic curvature \cite{KOIB-EPJB-2004,KOIB-PLA-2005-1}. It was reported that the model undergoes a first-order phase transition between the smooth phase and the crumpled phase on a sphere \cite{KOIB-EPJB-2004} and on a disk \cite{KOIB-PLA-2005-1}. As a consequence the phase structure of the model has been partly clarified: the transition can be seen independent of whether the surface is compact or non-compact. 

However, the model is not yet sufficiently understood. Remaining subject to be confirmed is whether the phase transition and the topology-change are {\it compatible} in the surface model on compact surfaces. Here {\it compatible} means that both of two phenomena lead to the same result without depending on which phenomenon firstly occurs. If we define the surface model on compact surfaces, it is reasonable to consider the topology-change of surfaces. In fact, the partition function for the closed string model of Polyakov includes the summation over topology. Moreover, a toroidal vesicle and a genus two vesicle with two holes can be observed in biological membranes \cite{David-TDQGRS-1989}.

Therefore, it is interesting, even in the context of the tethered surfaces, to study whether the first-order transition can be observed on a torus, as one of the higher genus surfaces. The phase transition might be seen only on spherical surfaces, if the phase transition and the topology-change are incompatible.  

In this Letter, in order to confirm the compatibility we will show that a first-order transition can be seen in the model on a torus with intrinsic curvature and that the phase transition is identical to that observed in the model on a sphere reported in \cite{KOIB-EPJB-2004}.

%------------------------------------------
\section{Model}\label{model}
%------------------------------------------
%++++++++++++++++++++++++++++++++++
\begin{figure}[htbp]
(a)\includegraphics[width=70mm]{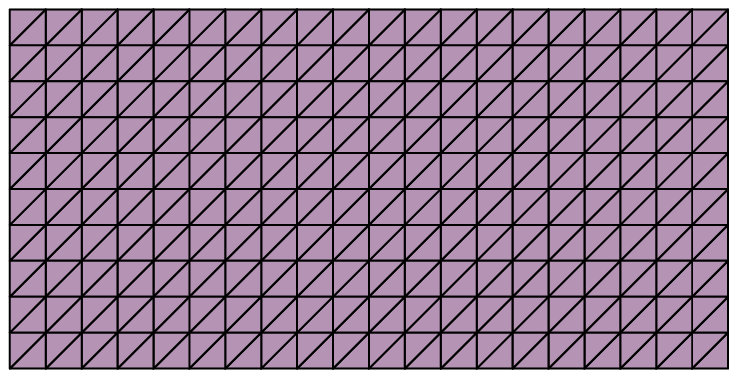}
(b)\includegraphics[width=70mm]{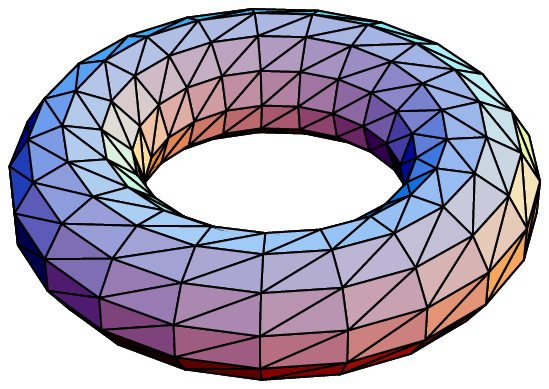}
  \caption{(a) Rectangular surface of size $L_1\!\times\!L_2\!=\!20\!\times\!10$, and (b) the real torus obtained by connecting the opposite sides of the surface in (a).}
  \label{fig-1}
\end{figure}
%++++++++++++++++++++++++++++++++++
A triangulated real torus is obtained by modifying a triangulated rectangular surface of size $L_1\times L_2$ as shown in Fig. \ref{fig-1}(a). Bending the surface and connecting the sides of length $L_1$, we have a cylinder of length $L_1$. Then the remaining two-sides of the cylinder are connected as in the first-step. Thus, we have a real torus as shown in Fig. \ref{fig-1}(b), which is topologically identical to the surface in Fig. \ref{fig-1}(a) under the periodic boundary condition. The real torus mentioned above will be used in the simulations.

The real torus is, therefore, characterized by the ratio $L_1/L_2$. Two kinds of tori are used in the simulations: $L_1/L_2\!=\!2$ and $L_1/L_2\!=\!4$. Figure \ref{fig-1}(b) is the torus of size $N\!=\!200$ of the first type $L_1/L_2\!=\!2$, where $L_1\!=\!20$ and $L_2\!=\!10$. Every vertex has a coordination number $\sigma\!=\!6$ on the torus.

The Gaussian tethering potential $S_1$ and the intrinsic curvature $S_3$ are defined by
\begin{equation}
\label{S3}
S_1=\sum_{(ij)} \left(X_i-X_j\right)^2, \; S_3= -\sum_i \log (\delta_i/2\pi), 
\end{equation}
where $\sum_{(ij)}$ is the sum over bond $(ij)$ connecting the vertices $X_i$ and $X_j$, and  $\delta_i$ in $S_3$ is the sum of angles of the triangles meeting at the vertex $i$, and $\sum_{i}$ is the sum over vertices $i$.

The partition function is defined by 
\begin{eqnarray}
 \label{partition-function}
Z(\alpha) = \int \prod _{i=1}^N dX_i \exp\left[-S(X)\right],\qquad \\
S(X)=S_1 + \alpha S_3,  \qquad\quad \nonumber
\end{eqnarray}
where $N$ is the total number of vertices, which is equal to $L_1\!\times\! L_2$ as described previously. The expression $S(X)$  shows that $S$ explicitly depends on the variable $X$. The coefficient $\alpha$ is a modulus of the intrinsic curvature. The surfaces are allowed to self-intersect, and the center of each surface is fixed in the partition function $Z(\alpha) $ to remove the translational zero-mode.

We note, as described in \cite{KOIB-EPJB-2004}, that the intrinsic curvature term  $S_3\!=\!-\sum_i \log (\delta_i/2\pi)$ comes from the integration measure $\prod_i dX_i q_i^\alpha$ \cite{DAVID-NPB-1985} in the partition function for the model on a sphere, where $q_i$ is the co-ordination number of the vertex $i$. The term $S_3\!=\!-\sum_i \log(\delta_i/2\pi)$ becomes minimum when $\delta_i\!=\!2\pi$ for all $i$, and hence it becomes smaller on a smooth torus than on a crumpled one. It is also exact that the term $S_3\!=\!\sum_i (\delta_i-2\pi)^2$ can be minimized on smooth configurations of the torus. The reason for using the term $S_3\!=\!-\sum_i \log (\delta_i/2\pi)$, as an intrinsic curvature on the torus as in the model on a sphere \cite{KOIB-EPJB-2004}, is that $S_3$ is closely related to the previously mentioned integration measure.

%------------------------------------------
\section{Monte Carlo technique}\label{MC-Techniques}
%------------------------------------------
We use two groups of surfaces classified by the ratio $L_1/L_2$ in the simulations as mentioned above. The first is characterized by $L_1/L_2\!=\!2$ and is composed of surfaces of size $N\!=\!1800$, $N\!=\!3200$, $N\!=\!5000$, and  $N\!=\!9800$. The second is characterized by $L_1/L_2\!=\!4$ and is composed of surfaces of size $N\!=\!1762$, $N\!=\!3600$, $N\!=\!6400$, and  $N\!=\!10000$.

The variables $X$ are updated by using the canonical Monte Carlo technique so that $X^\prime \!=\! X \!+\! \delta X$, where the small change $\delta X$ is made at random in a small sphere in ${\bf R}^3$. The radius $\delta r$ of the small sphere is chosen at the start of the simulations to maintain the rate of acceptance $r_X$ for the $X$-update as $0.4 \leq r_X \leq 0.6$.  

The total number of MCS (Monte Carlo sweeps) after the thermalization MCS is about $1.5\!\times\!10^8$ in the smooth phase at the transition point of surfaces of $N\!\geq\!5000$, and about $1.2\!\times\!10^8$ for the smaller surfaces. Relatively smaller number of MCS ($0.8\!\times\!10^8\sim 1.5\!\times\!10^8$) is iterated in the crumpled phase, because if the surfaces become once crumpled they hardly return smooth. This irreversibility was also seen in the model on a sphere \cite{KOIB-EPJB-2004} and in the model on a disk  \cite{KOIB-PLA-2005-1}. 

A random number called Mersenne Twister \cite{Matsumoto-Nishimura-1998} is used in the MC simulations. We use two sequences of random numbers; one for 3-dimensional move of vertices $X$ and the other for the Metropolis accept/reject in the update of $X$.

%------------------------------------------
\section{Results}\label{Results}
\begin{figure}[hbt]
\centering
\includegraphics[width=12cm]{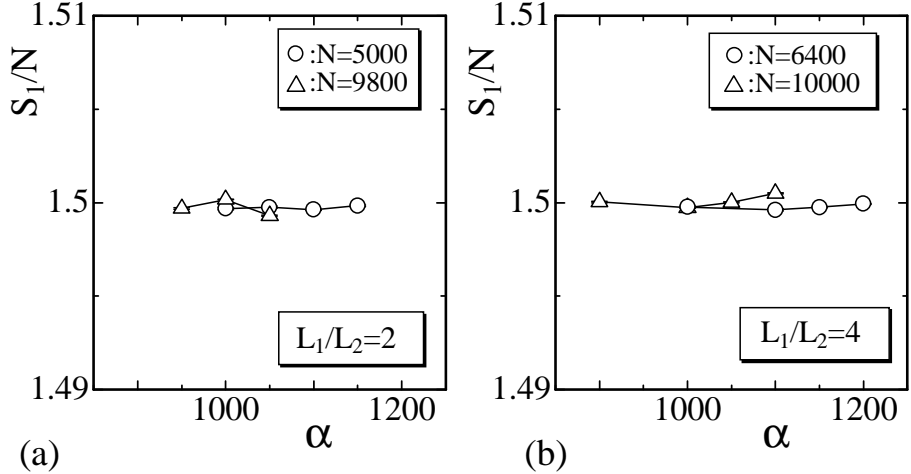}
\caption{$S_1/N$ vs. $\alpha$ obtained from the tori of types (a) $L_1/L_2\!=\!2$ and (b) $L_1/L_2\!=\!4$. }
\label{fig-2}
\end{figure}

%------------------------------------------
Figures \ref{fig-2}(a) and \ref{fig-2}(b) show $S_1/N$ against $\alpha$ obtained on the type $L_1/L_2\!=\!2$ surfaces and on the type $L_1/L_2\!=\!4$ surfaces, respectively. We find from these figures that the expected relation $S_1/N\!=\!1.5$ is satisfied. Scale invariance of the partition function predicts that $S_1/N\!=\!1.5$. We expect that this relation should not be influenced by whether the phase transition is of first order or not. 
%++++++++++++++++++++++++++++++++++
%++++++++++++++++++++++++++++++++++

\begin{figure}[hbt]
\centering
\includegraphics[width=12cm]{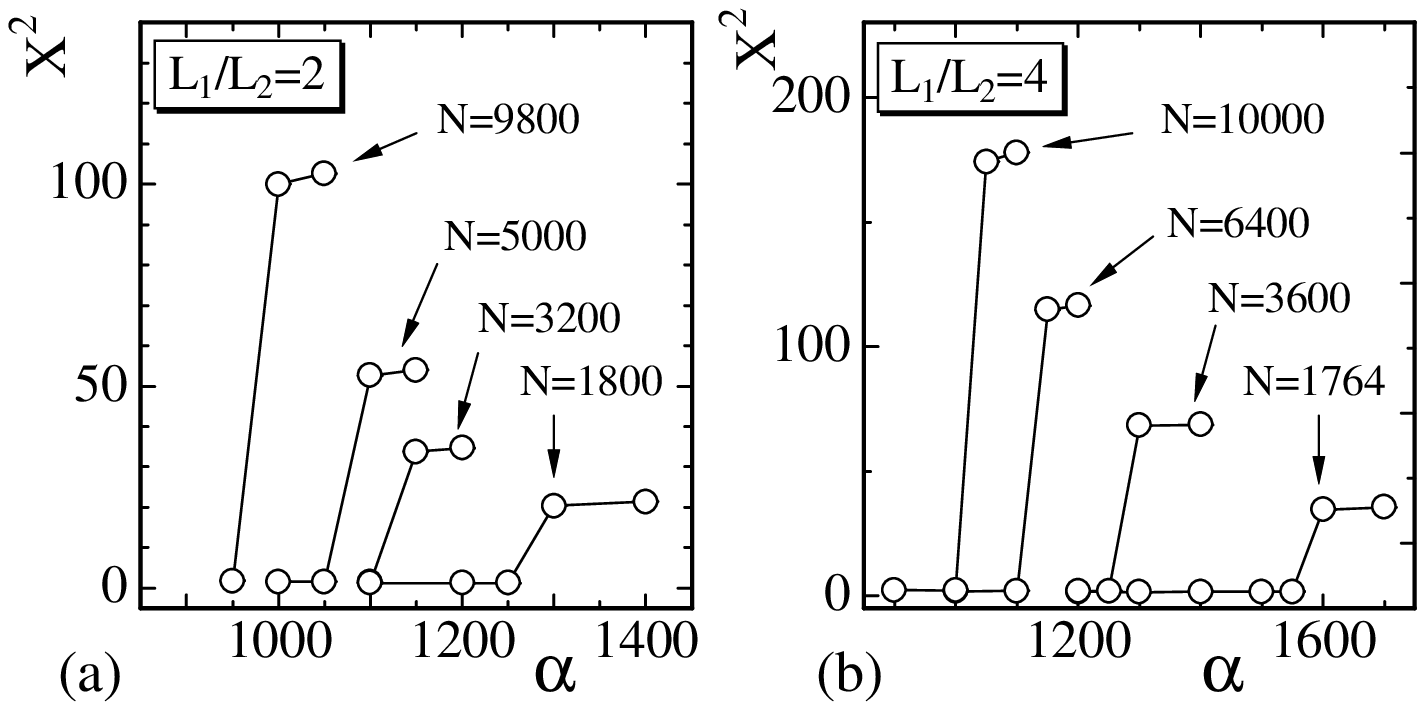}
\caption{$X^2$ vs. $\alpha$ obtained on surfaces of the type (a) $L_1/L_2\!=\!2$ and the type (b) $L_1/L_2\!=\!4$. The surfaces in (a) are of size from $N\!=\!1800$ to $N\!=\!9800$, and those in (b) are of size from $N\!=\!1764$ to $N\!=\!10000$.}
\label{fig-3}
\end{figure}

The size of surfaces can be reflected in the mean square size $X^2$ defined by
\begin{equation}
\label{X2}
X^2={1\over N} \sum_i \left(X_i-\bar X\right)^2, \quad \bar X={1\over N} \sum_i X_i.
\end{equation}
In fact, it is expected that the surfaces become smooth in the limit $\alpha\!\to\!\infty$ and crumpled in the limit $\alpha\!\to\!0$. In order to see how large the size of surfaces of the type $L_1/L_2\!=\!2$ is, we plot $X^2$ against $\alpha$ in Fig. \ref{fig-3}(a). Figure \ref{fig-3}(b) shows those for surfaces of the type $L_1/L_2\!=\!4$. We find a phase transition separating the smooth phase from the crumpled one both in Figs. \ref{fig-3}(a) and \ref{fig-3}(b). The transition seems as a first-order one because $X^2$ discontinuously changes at finite $\alpha$. 
%++++++++++++++++++++++++++++++++++
%++++++++++++++++++++++++++++++++++

%++++++++++++++++++++++++++++++++++
\begin{figure}[hbt]
\centering
\includegraphics[width=12cm]{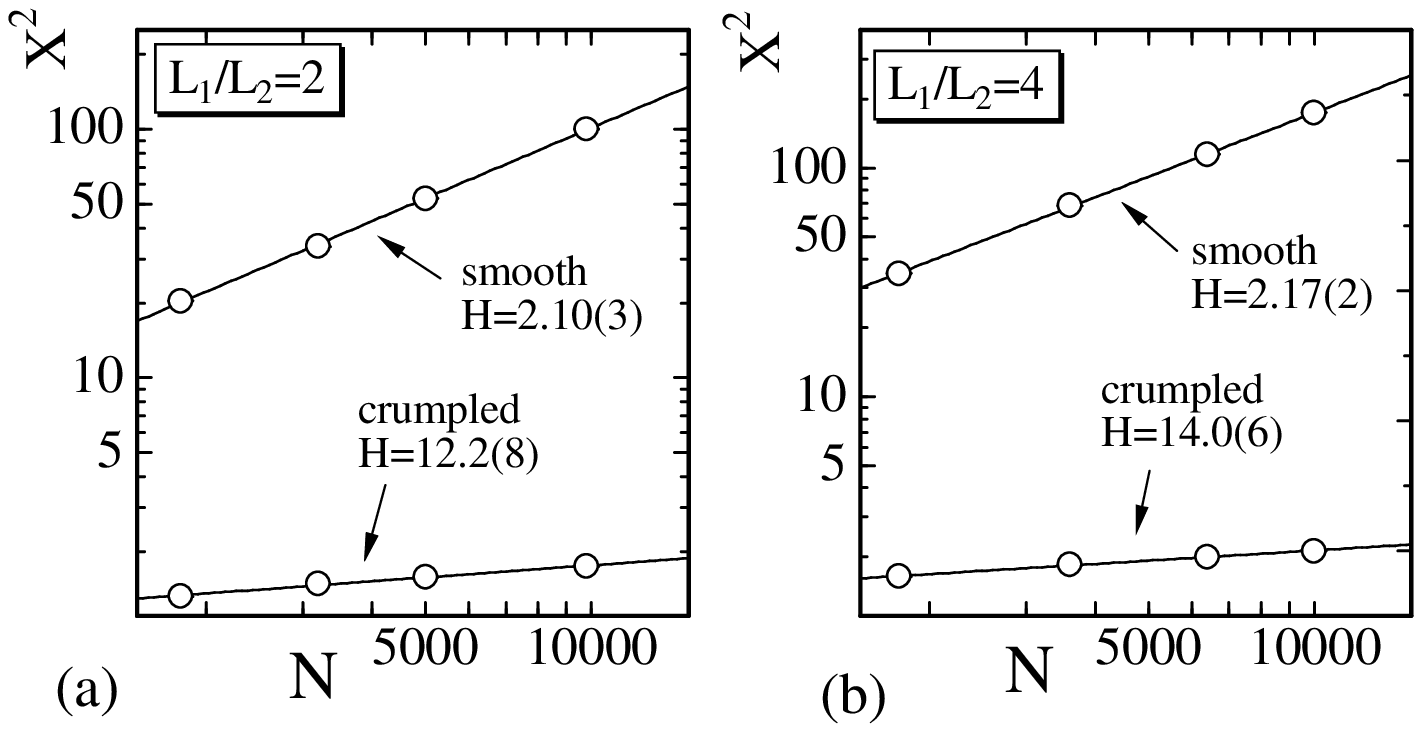}
\caption{ Log-log plots of $X^2$ vs. $N$ obtained just above and below the transition point $\alpha_c$ on surfaces of the types (a) $L_1/L_2\!=\!4$ and (b) $L_1/L_2\!=\!4$. The straight lines in (a) and (b) are drawn by fitting the data to Eq. (\ref{Hausdorff}).}
\label{fig-4}
\end{figure}
%++++++++++++++++++++++++++++++++++
Figures \ref{fig-4}(a) and \ref{fig-4}(b) show log-log plots of $X^2$ against $N$, obtained just below and above the transition point $\alpha_c$ where $X^2$ discontinuously changes as shown in Figs. \ref{fig-3}(a) and \ref{fig-3}(b). The straight line denoted by smooth (crumpled) is obtained by fitting $X^2$ in the smooth (crumpled) phase just above (below) $\alpha_c$ by
\begin{equation}
\label{Hausdorff}
X^2 \sim N^{2/H},
\end{equation}
where $H$ is the Hausdorff dimension. Thus we have 
\begin{equation}
\label{H1}
H(L_1/L_2\!=\!2)=\left\{ \begin{array}{ll}
    2.10\pm 0.03,                   & (\mbox{smooth}); \\
   12.2\pm 0.7, & (\mbox{crumpled}), 
   \end{array} \right.
\end{equation}
\begin{equation}
\label{H2}
H(L_1/L_2\!=\!4)=\left\{ \begin{array}{ll}
    2.17\pm 0.02,                   & (\mbox{smooth}); \\
   14.0\pm 0.6, & (\mbox{crumpled}). 
   \end{array} \right.
\end{equation}
The results $H\!=\!2.10(3)$ in (\ref{H1}) and $H\!=\!2.17(2)$ in (\ref{H2})  imply that the surfaces are almost smooth in the smooth phase. On the contrary,  $H\!=\!12.2(7)$ in (\ref{H1}) and $H\!=\!14.0(6)$ in (\ref{H2})  imply that the surfaces are highly crumpled in the crumpled phase close to the transition point.

%++++++++++++++++++++++++++++++++++
\begin{figure}[hbt]
\centering
\includegraphics[width=12cm]{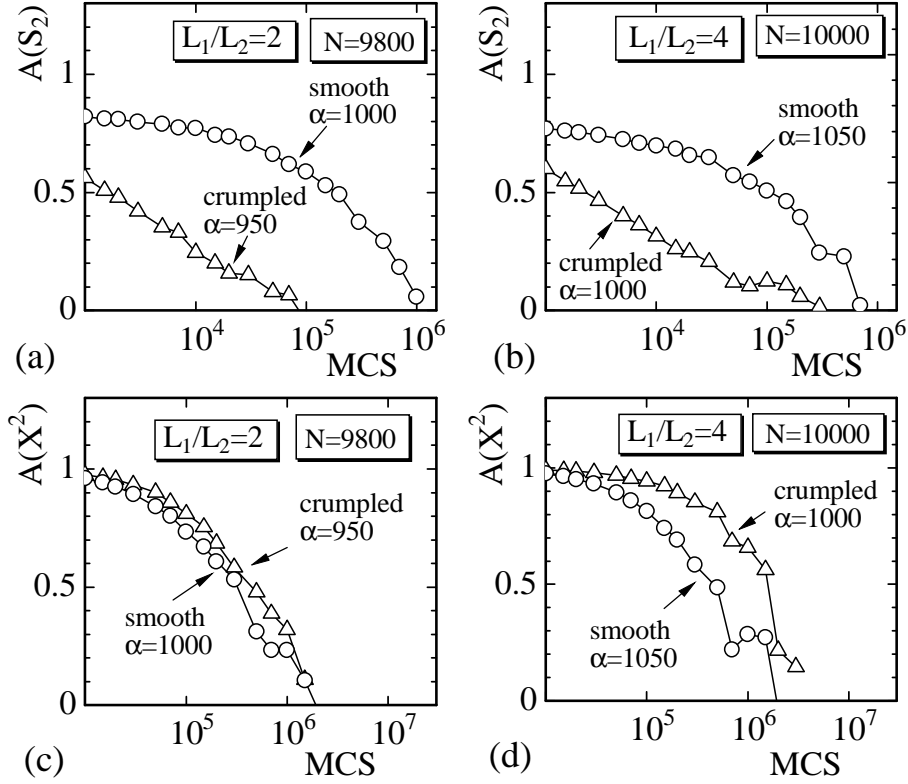}
\caption{The auto-correlation coefficient $A(S_2)$ in the smooth and the crumpled phases on the tori of the types (a) $L_1/L_2\!=\!2$ and (b) $L_1/L_2\!=\!4$. The auto-correlation coefficient $A(X^2)$  obtained on the tori of the types (c) $L_1/L_2\!=\!2$ and (d) $L_1/L_2\!=\!4$. }
\label{fig-5}
\end{figure}
%++++++++++++++++++++++++++++++++++

The phase transition can also be reflected in the convergence speed of the MC simulations. In order to see the convergence speed, we plot the autocorrelation coefficient $A(S_2)$ in Figs. \ref{fig-5}(a) and \ref{fig-5}(b). The autocorrelation coefficient is defined by 
\begin{eqnarray}
A(S_2)= \frac{\sum_i S_2(\tau_{i}) S_2(\tau_{i+1})} 
   {  \left[\sum_i  S_2(\tau_i)\right]^2 },\\ \nonumber
 \tau_{i+1} = \tau_i + n \times 500, \quad n=1,2,\cdots,  
\end{eqnarray}
where $S_2$ is the bending energy defined by $S_2\!=\!\sum_i(1-\cos\theta_i)$. The horizontal axes in the figures represent $500\!\times\! n\;(n\!=\!1,2,\cdots)$-MCS, which is a sampling-sweep between the samples $S_2(\tau_i)$ and $S_2(\tau_{i+1})$. The coefficients $A(S_2)$ in Figs. \ref{fig-5}(a) and \ref{fig-5}(b) are obtained on surfaces of $L_1/L_2\!=\!2$ and $L_1/L_2\!=\!4$ respectively. From these figures, we can find that the convergence speed for $S_2$ in the smooth phase is 10 times or more larger than that in the crumpled phase. 

Figures \ref{fig-5}(c) and \ref{fig-5}(d) show the autocorrelation coefficient $A(X^2)$ for the mean square size $X^2$. Contrary to the behavior of $A(S_2)$, there is no difference in $A(X^2)$ between the crumpled phase and the smooth phase. The reason why the convergence speed for $X^2$ in the smooth phase is almost identical to the one in the crumpled phase is that the phase space volume ($\subseteq {\bf R}^3$), where the vertices $X_i$ take their values, is almost the same both in the smooth phase and in the crumpled phase. The shape of surfaces in the smooth phase is considered to be almost fixed, although they extend to relatively large region in ${\bf R}^3$. This behavior of $A(X^2)$ is very similar to the one in a model on a sphere with extrinsic curvature \cite{KOIB-PRE-2005-1} but very different from the one in the model on a disk with intrinsic curvature \cite{KOIB-PLA-2005-1}, where $A(X^2)$ in the smooth phase is quite larger than that in the crumpled phase close to the transition point. Therefore, we understand from $A(S_2)$ and $A(X^2)$ in Figs. \ref{fig-5}(a) -- \ref{fig-5}(d) that the toroidal surface can fluctuate only locally, and there is no large-deformation of the surface in the smooth phase. It is quite likely that the shape of surface in the smooth phase is almost unchanged in the model on compact surfaces, and it largely changes in the model on non-compact surfaces so that the phase space volume may be relatively large.     
 
The phenomenon of critical slowing down can not be seen both in $A(S_2)$ and in $A(X^2)$ at the transition point. The reason of this is because of irreversibility of the transition. The transition from the crumpled phase to the smooth one appears to be irreversible. As mentioned in the chapter \ref{MC-Techniques}, if the vertices once localize to a small region in ${\bf R}^3$ at the transition point, they hardly expand to be a smooth surface.

%++++++++++++++++++++++++++++++++++
\begin{figure}[hbt]
\centering
\includegraphics[width=12cm]{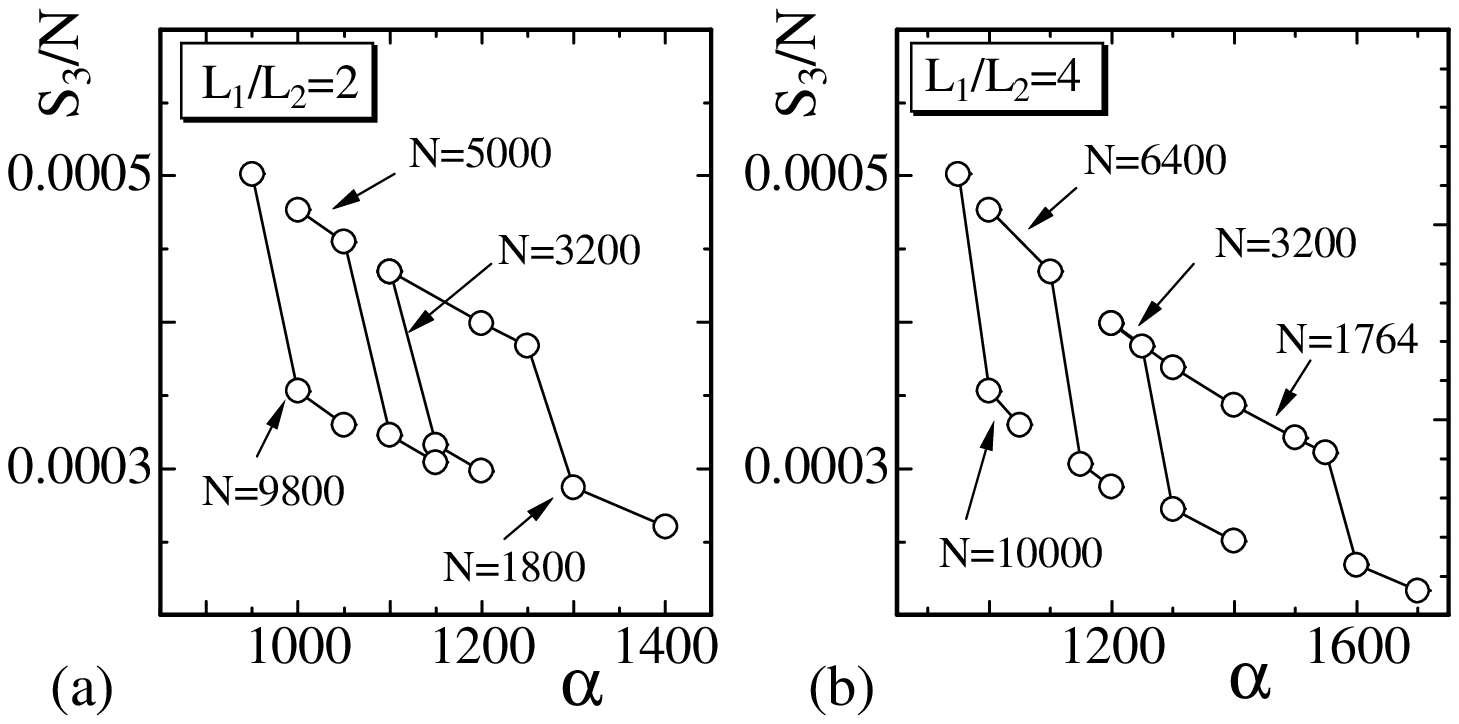}
\caption{$S_3/N$ vs. $\alpha$ obtained on the tori of (a) $L_1/L_2\!=\!4$ and (b) $L_1/L_2\!=\!4$.  }
\label{fig-6}
\end{figure}
%++++++++++++++++++++++++++++++++++
%++++++++++++++++++++++++++++++++++
\begin{figure}[hbt]
\centering
\includegraphics[width=12cm]{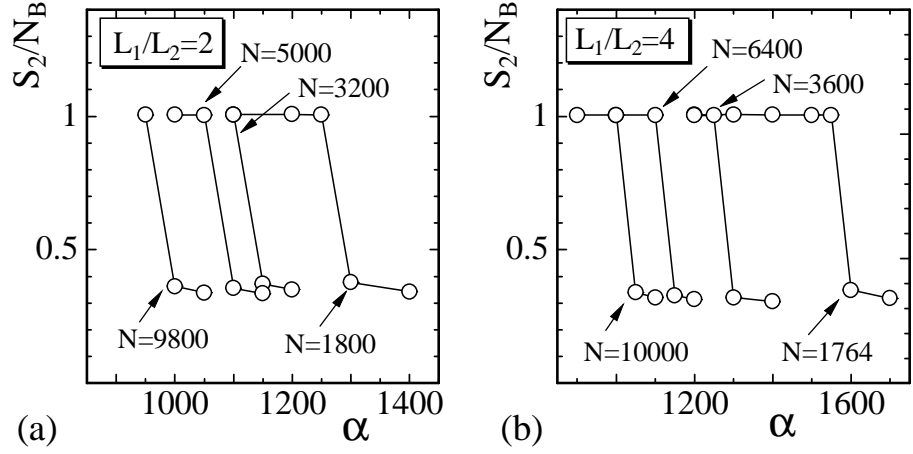}
\caption{$S_2/N_B$ vs. $\alpha$ on the tori of (a) $L_1/L_2\!=\!4$ and (b) $L_1/L_2\!=\!4$, where $N_B(=\!3N)$ is the total number of bonds.}
\label{fig-7}
\end{figure}
%++++++++++++++++++++++++++++++++++
The intrinsic curvature energy $S_3/N$ is shown in Figs. \ref{fig-6}(a) and \ref{fig-6}(b), and it is clear that $S_3/N$ changes discontinuously against $\alpha$. This indicates the existence of a first-order phase transition. The variation of $S_2/N_B$ against $\alpha$ is shown in Figs. \ref{fig-7}(a) and \ref{fig-7}(b), where $N_B$ is the total number of bonds, which is given by $N_B\!=\!3N$. The discontinuous change of $S_2/N_B$ indicates that the phase transition separates the smooth phase from the crumpled phase. 

%++++++++++++++++++++++++++++++++++
\begin{figure}[hbt]
\centering
\includegraphics[width=12cm]{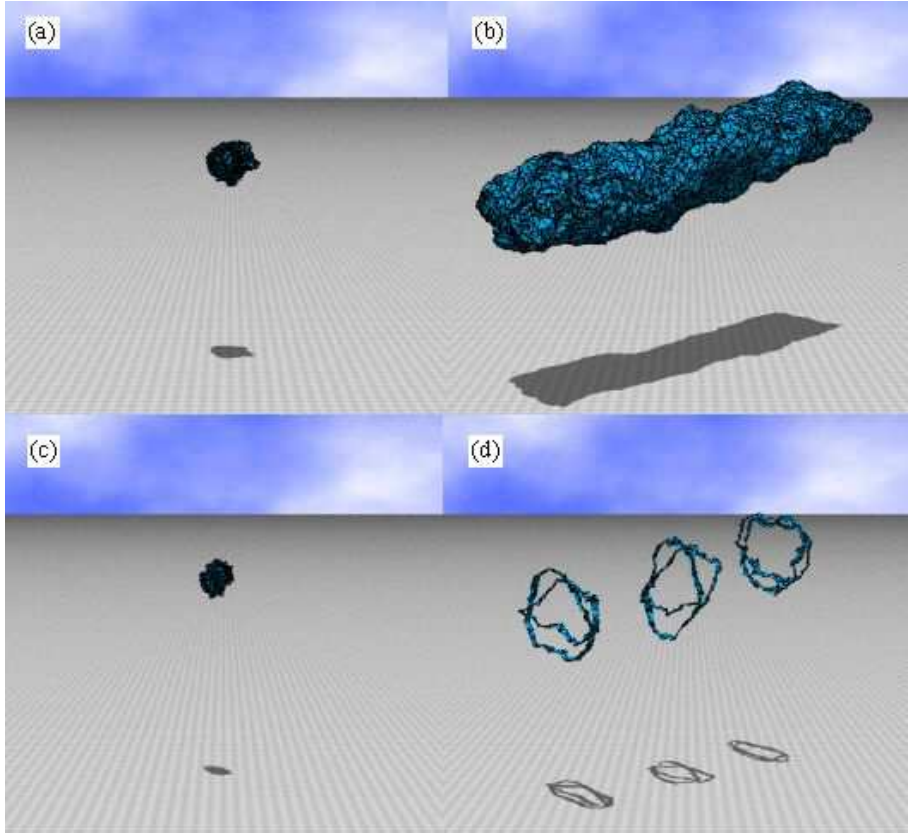}
\caption{Snapshots of surfaces of the type $L_1/L_2\!=\!4$ and of size $N\!=\!10000$ at (a) $\alpha\!=\!1000$, and at (b) $\alpha\!=\!1050$, and (c) a section of the surface in (a), (d) sections of the surface in (b).  Figures are shown in the same scale. }
\label{fig-8}
\end{figure}
%++++++++++++++++++++++++++++++++++
Snapshots of surfaces of the type $L_1/L_2\!=\!4$ are shown in Figs. \ref{fig-8}(a) and \ref{fig-8}(b), which are obtained at $\alpha\!=\!1000$ and $\alpha\!=\!1050$. The size of surfaces in Figs. \ref{fig-8}(a) and \ref{fig-8}(b) is of $N\!=\!10000$.  Sections of the surfaces in Figs. \ref{fig-8}(a) and \ref{fig-8}(b) are shown in Figs. \ref{fig-8}(c) and \ref{fig-8}(d) respectively. It is easy to see that the surface in Fig. \ref{fig-8}(a) is crumpled and that the surface in Fig. \ref{fig-8}(b) is smooth. Although the surface in Fig. \ref{fig-8}(b) seems to be a cylinder or a tube, it is almost smooth as can be seen in the section shown in Fig. \ref{fig-8}(d).  We note that the cylindrical surface in Fig. \ref{fig-8}(b) is rotationally symmetric only around the longitudinal axis of the surface. 

The appearance of an axis along such cylindrical and smooth surface in Fig. \ref{fig-8}(b) does not imply a spontaneous symmetry breaking, which can be seen in a surface model in \cite{KOIB-PRE-2004-2}. In fact, a real torus such as the one shown in Fig. \ref{fig-1}(b) is also rotationally symmetric only along the axis perpendicular to the hole of the torus. In this sense, a symmetry of the model in the smooth phase is equivalent to the one of the real torus shown in Fig. \ref{fig-1}(b). On the contrary, the surface in the crumpled phase is expected to have a higher symmetry just like in the crumpled phase of the model on a sphere and on a disk. 

%------------------------------------------
\section{Summary and conclusion}\label{Conclusions}
%------------------------------------------
In summary, our attentions were focused on the compatibility of the topology change and the phase transition in a tethered surface model on compact surfaces with intrinsic curvature. 
In order to see this compatibility, we have studied the model on two kinds of tori in ${\bf R}^3$; one is characterized by $L_1/L_2\!=\!2$ and the other by $L_1/L_2\!=\!4$, where $L_1$ and $L_2$ are the length of two sides of rectangles which are deformed to be the tori by connecting the sides. Monte Carlo simulations were performed on surfaces of size up to $N(\!=\!L_1\!\times\!L_2)\!\simeq\!1\!\times\!10^4$ in both kinds of tori. 

We found that the model undergoes a first-order phase transition separating the smooth phase from the crumpled one on both kinds of tori. The smooth phase is characterized by Hausdorff dimension $H\!\simeq\!2$, which is identical with the topological dimension of surfaces. The surface in the crumpled phase is completely collapsed, and characterized by large values of Hausdorff dimension. These results indicate that the phase transition on a torus is identical with that on a sphere \cite{KOIB-EPJB-2004} and that on a disk \cite{KOIB-PLA-2005-1}. Thus we can conclude that the topology change and the phase transition is compatible in the tethered model of surfaces with the intrinsic curvature. Moreover, the autocorrelation coefficients $A(X^2)$ obtained on the tori in this paper and on the disk in \cite{KOIB-PLA-2005-1} enable us to consider that compact surfaces deform very little in the smooth phase, and on the contrary, non-compact surfaces can largely deform in the smooth phase. 

An interesting problem to be studied is whether the compatibility is present in the model with extrinsic curvature, which exhibits a first-order transition on a sphere \cite{KOIB-EPJB-2004}. It is also interesting to see how the phase transition of the intrinsic curvature model is influenced by the fluidity of lateral diffusion, which is realized on dynamically triangulated surfaces.

This work is supported in part by a Grant-in-Aid for Scientific Research, No. 15560160.

%----------------------------------------------------------
%\vspace*{3mm}
%\noindent
%{\bf Acknowledgment}\\
%\vspace*{2mm}
%\par

%\vfill\eject


\vspace*{5mm}
\noindent
\begin{thebibliography}{999}
%---------------------------
%----------------------------
\bibitem{HELFRICH-NF-1973}
 W.Helfrich, Z. Naturforsch \textbf{28c}, (1973) 693.

\bibitem{POLYAKOV-NPB-1986}
 A.M. Polyakov, Nucl. Phys. B \textbf{268}, (1986) 406.

\bibitem{Kleinert-PLB-1986}
 H. Kleinert, Phys. Lett. B \textbf{174}, (1986) 335.

\bibitem{David-TDQGRS-1989}
F. David,  in {Two dimensional quantum gravity and random surfaces, Vol.8}, edited by  D. Nelson, T. Piran, and S. Weinberg, (World Scientific, Singapore, 1989), p.81.

\bibitem{NELSON-SMMS-2004}
 D. Nelson, \textit{Statistical Mechanics of Membranes and Surfaces, Second Edition},  
 D. Nelson, T. Piran, and S. Weinberg  Eds. (World Scientific, 2004) p.1.


%----------------------------
\bibitem{DavidGuitter-EPL-1988}
 F. David, and  E.Guitter, Europhys. Lett \textbf{5(8)}, (1988) 709.

\bibitem{Peliti-Leibler-PRL-1985}
 L.Peliti, and S. Leibler, Phys. Rev. Lett. \textbf{54},  (1985) 1690.

\bibitem{BKS-PLA-2000}
 M.E.S. Borelli, H. Kleinert, and  A.M.J Schakel, 
Phys.Lett.A \textbf{267}, (2000) 201.

\bibitem{BK-PRB-2001}
 M.E.S. Borelli, and H. Kleinert, 
Phys. Rev. B 63, (2001) 205414. 

\bibitem{Kleinert-EPJB-1999}
 H. Kleinert, Euro. Phys. J. B \textbf{9}, (1999) 651.

\bibitem{BOWICK-TRAVESSET-PREP-2001}
 M.J. Bowick, and  A. Travesset, Phys. Rept. \textbf{344}, (2001) 255.

\bibitem{WIESE-PTCP19-2000}
 K.J. Wiese, 
\textit{Phase Transitions and Critical Phenomena 19}, 
 C. Domb, and  J.L. Lebowitz Eds. (Academic Press, 2000) p.253.

\bibitem{WHEATER-JP-1994}
 J.F. Wheater, J. Phys. A:Math.Gen \textbf{27}, (1994) 3323.

%----------------------------

\bibitem{CATTERALL-NPB-SUPL-1991}
 S.M. Catterall,  J.B. Kogut, and  R.L. Renken, 
Nucl. Phys. Proc. Suppl. B \textbf{99A}, (1991) 1.

\bibitem{AMBJORN-NPB-1993}
 J. Ambjorn,  A. Irback,  J. Jurkiewicz, and  B. Petersson, 
Nucl. Phys. B \textbf{393}, (1993) 571.

\bibitem{BCHGM-NPB-1993}
M. Bowick,  P. Coddington,  L. Han, G. Harris, and  E. Marinari, 
 Nucl. Phys. Proc. Suppl. \textbf{30}, (1993) 795;
 Nucl. Phys. B \textbf{394}, (1993) 791.

\bibitem{ABGFHHE-PLB-1993}
 K. Anagnostopoulos,  M. Bowick,  P. Gottington,  M. Falcioni,
  L. Han,  G. Harris, and  E. Marinari, Phys. Lett. B \textbf{317}, (1993) 102.

\bibitem{KOIB-PLA-2002}
 H. Koibuchi, Phys. Lett. A \textbf{300}, (2002) 586.

\bibitem{KOIB-PLA-2004}
 H. Koibuchi,  N. Kusano,  A. Nidaira, and K. Suzuki, 
Phys. Lett. A  \textbf{332}, (2004) 141. 

\bibitem{KOIB-EPJB-2005}
 H. Koibuchi, 
Eur. Phys. J. B \textbf{45}, (2005) 377. 

%----------------------------
\bibitem{KOIB-PLA-2003-2}
 H. Koibuchi,  N. Kusano,  A.Nidaira,  K.Suzuki, and  M. Yamada, 
Phys. Lett. A  \textbf{319}, (2003) 44. 

\bibitem{KOIB-PRE-2003}
 H. Koibuchi, A. Nidaira, T. Morita, and K. Suzuki,
Phys. Rev. E  \textbf{68}, (2003) 011804. 

\bibitem{KOIB-EPJB-2004}
H.Koibuchi, N.Kusano, A.Nidaira, Z.Sasaki and K.Suzuki,
Eur. Phys. J. B \textbf{42}, (2004) 561.

\bibitem{KOIB-PLA-2005-1}
 M. Igawa, H. Koibuchi,  and M.Yamada, 
Phys. Lett. A  \textbf{338}, (2005) 338. 


%----------------------------


\bibitem{BCFTA-1996-1997}
 M. Bowick,  S. Catterall,  M. Falcioni,  G. Thorleifsson, and  K. Anagnostopoulos, 
J. Phys. I France \textbf{6}, (1996) 1321; 
Nucl. Phys. Proc. Suppl. \textbf{47}, (1996) 838; 
Nucl. Phys. Proc. Suppl. \textbf{53}, (1997) 746.

\bibitem{WHEATER-NPB-1996}
 J.F. Wheater, Nucl. Phys. B \textbf{458}, (1996) 671.

\bibitem{KANTOR-NELSON-PRA-1987}
 Y. Kantor, and  D.R. Nelson, Phys. Rev. A \textbf{36}, (1987) 4020.

\bibitem{GOMPPER-KROLL-PRE-1995}
 G. Gompper, and  D.M. Kroll, Phys. Rev. E \textbf{51}, (1995) 514.

\bibitem{KOIB-PRE-2004-1}
 H. Koibuchi,  N. Kusano,  A. Nidaira, K. Suzuki, and M. Yamada, 
Phys. Rev. E  \textbf{69}, (2004) 066139. 

\bibitem{KOIB-PRE-2004-2}
 H. Koibuchi,  Z. Sasaki, and K. Shinohara, 
Phys. Rev. E  \textbf{70}, (2004) 066144. 

\bibitem{KOIB-PRE-2005-1}
 H. Koibuchi,  and T. Kuwahata, 
Phys. Rev. E \textbf{72}, (2005) 026124.


%----------------------------
\bibitem{KANTOR-SMMS-2004}
 Y. Kantor, 
\textit{Statistical Mechanics of Membranes and Surfaces, Second Edition},
 D. Nelson,  T. Piran, and  S. Weinberg Eds. (World Scientific, 2004) p.111.

\bibitem{KANTOR-KARDER-NELSON-PRA-1987}
 Y. Kantor, M. Karder, and  D.R. Nelson, Phys. Rev. A \textbf{35}, (1987) 3056.

\bibitem{BCTT-EPJE-2001}
 M. Bowick,  A. Cacciuto,  G.Thorleifsson, and  A. Travesset, 
Eur. Phys. J. E \textbf{5}, (2001) 149. 

\bibitem{BOWICK-SMMS-2004}
 M. J. Bowick, 
\textit{Statistical Mechanics of Membranes and Surfaces, Second Edition},
 D. Nelson,  T. Piran, and  S. Weinberg Eds. (World Scientific, 2004) p.323.

%----------------------------
\bibitem{NELSON-SMMS-2004-2}
 D. Nelson, \textit{Statistical Mechanics of Membranes and Surfaces, Second Edition},  
 D. Nelson, T. Piran, and S. Weinberg  Eds. (World Scientific, 2004) p.131.

\bibitem{FDAVID-SMMS-2004-2}
 F. David, \textit{Statistical Mechanics of Membranes and Surfaces, Second Edition},  
 D. Nelson, T. Piran, and S. Weinberg  Eds. (World Scientific, 2004) p.149.



%----------------------------
\bibitem{BJ-PRD-1993-1994}
C.F. Baillie, and D.A. Johnston,  Phys. Rev. D {\bf48}, (1993)  5025; {\bf 49}, (1994)4139.

\bibitem{BEJ-PLB-1993}
C.F. Baillie, D.Espriu, and D.A. Johnston, Phys. Lett. {\bf 305B}, (1993) 109.

\bibitem{BIJJ-PLB-1994}
C.F. Baillie, A.Irback, W.Janke and D.A.Johnston, Phys. Lett. {\bf 325B}, (1994) 45.

\bibitem{FW-PLB-1993}
N.Ferguson, and J.F.Wheater, Phys. Lett. {\bf 319B}, (1993) 104.

%----------------------------

%----------------------------
\bibitem{DAVID-NPB-1985}
 F. David,
 Nucl. Phys. B \textbf{257[FS14]} (1985) 543.

%---------------------------
\bibitem{Matsumoto-Nishimura-1998}
M. Matsumoto and T. Nishimura, "Mersenne Twister: A 623-dimensionally equidistributed uniform  pseudorandom number generator", ACM Trans. on Modeling and Computer Simulation Vol. 8, No. 1,  January pp.3-30 (1998).
%----------------------------


\end{thebibliography}
\end{document}